\documentclass[sigconf,screen]{acmart}
\usepackage{algorithm}
\usepackage{algpseudocode}
\usepackage{caption}

\newcommand{\ignore}[1]{}
\newcommand{\vd}[1]{ 
	\textcolor{blue}{(VD says: #1)}}

\newcommand{\ts}[1]{ 
	\textcolor{teal}{(Tanay says: #1)}}  

\AtBeginDocument{%
  }

\setcopyright{acmlicensed}
\copyrightyear{}
\acmYear{}
\acmDOI{}
\acmSubmissionID{370}
\acmConference[]{}{}{}

\begin{document}

\title[Reinforcement Learning–Based Co-Design and Operation for Cost-Optimal HVAC Systems]{Reinforcement Learning–Based Co-Design and Operation of Chiller and Thermal Energy Storage for Cost-Optimal HVAC Systems}
\author{Tanay Raghunandan Srinivasa}
\affiliation{%
  \institution{Plaksha University}
  \city{Mohali}
  \country{India}
}

\email{tanay.srinivasa@plaksha.edu.in}

\author{Vivek Deulkar}
\affiliation{%
  \institution{Plaksha University}
  \city{Mohali}
  \country{India}
}
\email{vivek.deulkar@plaksha.edu.in}

\author{Aviruch Bhatia}
\affiliation{%
  \institution{Plaksha University}
  \city{Mohali}
  \country{India}
}
\email{aviruch.bhatia@plaksha.edu.in}

\author{Vishal Garg}
\affiliation{%
  \institution{Plaksha University}
  \city{Mohali}
  \country{India}
}
\email{vishal.garg@plaksha.edu.in}

\renewcommand{\shortauthors}{}


\begin{abstract}

We study the joint operation and sizing of cooling infrastructure for commercial HVAC systems using reinforcement learning, with the objective of minimizing life-cycle cost over a 30-year horizon. The cooling system consists of a fixed-capacity electric chiller and a thermal energy storage (TES) unit, jointly operated to meet stochastic hourly cooling demands under time-varying electricity prices. The life-cycle cost accounts for both capital expenditure and discounted operating cost, including electricity consumption and maintenance.

A key challenge arises from the strong asymmetry in capital costs: increasing chiller capacity by one unit is far more expensive than an equivalent increase in TES capacity. As a result, identifying the right combination of chiller and TES sizes, while ensuring zero loss-of-cooling-load under optimal operation, is a non-trivial co-design problem. To address this, we formulate the chiller operation problem for a fixed infrastructure configuration as a finite-horizon Markov Decision Process (MDP), in which the control action is the chiller part-load ratio (PLR). The MDP is solved using a Deep Q Network (DQN) with a constrained action space.

The learned DQN policy minimizes electricity cost over historical traces of cooling demand and electricity prices, while explicitly accounting for chiller inefficiencies as functions of temperature and part-load ratio, and for TES charging and discharging losses. For each candidate chiller–TES sizing configuration, the trained policy is evaluated to obtain the resulting operational cost and any unmet cooling load. We then restrict attention to configurations that fully satisfy the cooling demand and perform a life-cycle cost minimization over this feasible set to identify the cost-optimal infrastructure design. Using this approach, we determine the optimal chiller and thermal energy storage capacities to be 700 and 1500, respectively.

 \end{abstract}
%

\ccsdesc[500]{Computing methodologies~Reinforcement learning}
\ccsdesc[500]{Applied computing~Economics}
\ccsdesc[500]{Applied computing~Decision analysis}

\keywords{HVAC systems, Thermal energy storage, Building cooling demand, Reinforcement learning, Deep Q-Networks (DQN)}


\maketitle

\section{Introduction}
\begin{figure}[t]
    \centering
    \includegraphics[width=0.9\linewidth, trim=8pt 7pt 12pt 1pt, clip]{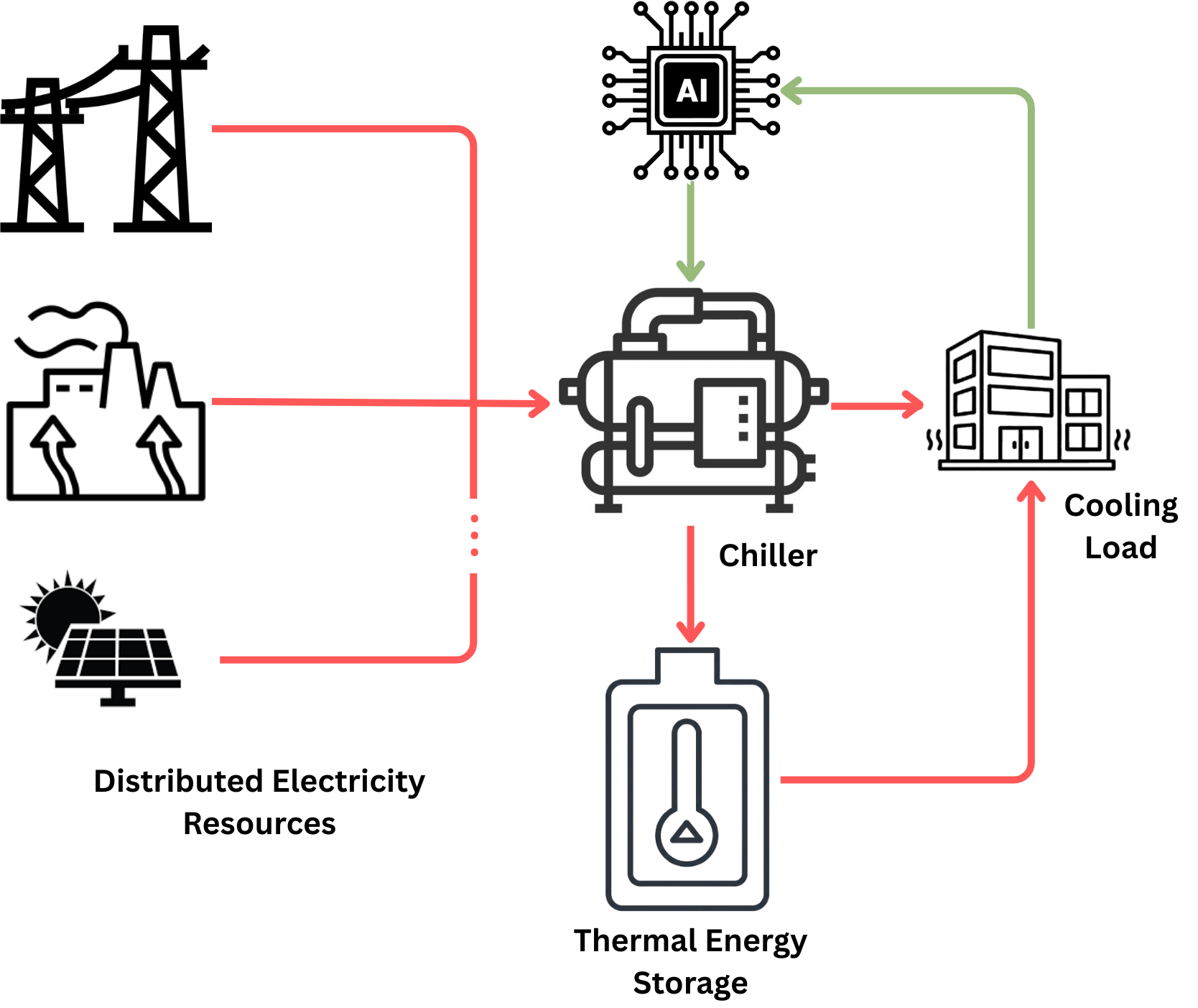}
    \caption{Illustration of Distributed HVAC System}
    \label{fig:introductory-Figure}
    \captionsetup[figure]{skip=0pt}
\end{figure}

Heating, Ventilation, and Air Conditioning (HVAC) systems are major consumers of electricity in commercial buildings \cite{GONZALEZTORRES2022626}. Cooling demand varies diurnally and seasonally, while electricity prices fluctuate due to market conditions and resource availability. Conventional designs often size chillers to meet peak cooling demand and operate them myopically to satisfy instantaneous loads, leading to high capital and operating costs.


Thermal energy storage (TES) provides a mechanism to shift cooling production in time: a chiller can produce extra cooling when electricity is cheap, charge TES, and later discharge TES to meet load when electricity is expensive \cite{bruce2019}.
Thermal Energy Storage (TES) provides temporal flexibility by decoupling cooling production from consumption \cite{HEIER20151305}. With the increasing penetration of renewable energy in the grid, the use of thermal energy storage (TES) becomes increasingly important \cite{VEREZ2023106344}. 
Excess cooling generated during low-price periods can be stored and later discharged to meet demand during high-price periods. However, the economic benefit of TES depends critically on how the chiller is operated over time and how the chiller and TES are jointly sized. Oversizing the chiller increases capital expenditure disproportionately relative to TES, while undersizing risks unmet cooling load. Controlling TES systems is important to get optimised performance \cite{YU2015203} and EMS can help in cost reduction \cite{app15020880}. 

Previous studies have applied hybrid machine-learning approaches to predict the performance of air-conditioning systems integrated with thermal energy storage (TES), addressing the limitations of conventional modeling methods. One such approach combines a radial basis function neural network with a meta-heuristic optimizer to capture nonlinear system behavior and improve prediction accuracy using in-situ measurement data. The results demonstrate superior predictive performance compared to existing HVAC-focused machine-learning models \cite{IRSHAD2024111308}. A review paper from the University of Lleida \cite{MEHRAJ2025116870} presents a comprehensive analysis of AI-based techniques for the design and optimization of thermal energy storage (TES) systems, benchmarking them against conventional approaches, highlighting their advantages and limitations, and examining their applications within TES systems. Mehraj N. \cite{MEHRAJ2026129452} used genetic algorithm based methods for optimal thermal energy storage tank design and demonstrated how computational intelligence integrated with physical principles yielded high-performance thermal energy storage systems. Bastida H. \cite{BASTIDA2024123526} used discrete-time state-of-charge estimator for latent heat thermal energy storage units based on a recurrent neural network and found that the key advantage of the RNN-based SoC estimator lies in its computational efficiency. It operates with a reduced set of matrix operations and activation functions, resulting in a decreased computation time (24 times faster) when compared to a discrete-time non-linear observer
based SoC estimator. Odufuwa O. Y. \cite{ODUFUWA2024112547} used ANN for predicting perfomance of ice thermal enregy stroage and multiple parameters storage temperature, COP, cooling load, and chiller power consumption are covered in the analysis. 
Lee D. \cite{LEE2022103700} used AI to control the TES system and validated it by the experimental analysis and showed that AI-based MPC was highly feasible by flexibly managing the different load levels and reducing the operating cost by 9.1–14.6\%.

Reinforcement learning has also been widely applied in the control of HVAC systems \cite{Ozan} and Thermal Energy Storage. Study done by Gregor P et al. \cite{Gregor01072003} investigated reinforcement learning control for operation of electrically driven cool thermal energy storage systems in commercial buildings, where the reinforcement learning controller learns to charge and discharge a thermal storage tank based on the feedback it receives from past control actions. Wang X. et al. \cite{WANG2023113696} proposed the use of a reinforcement learning (RL) framework for a commercial building. When compared to the fixed-schedule strategy, the RL controller resulted in a 7.6\% cost reduction during the simulated cooling season of 2020. This RL-based control method is capable of effectively learning system characteristics and enhancing the cost efficiency of the ice-based thermal energy storage (TES) system.
Study by Jiang Z. et al. \cite{JIANG2021110833} developed Deep Q-Network with an action processor, defining the environment as a Partially Observable Markov Decision Process with a reward function consisting of energy cost (time-of-use and peak demand charges) and a discomfort penalty. Zeyang Li et al. \cite{LI2023108742} worked on DR strategies based on RL, active thermal energy storage, and time-of-use electricity prices are formulated to find the optimal indoor T and RH setpoints, considering environmental constraints, comfort levels, and energy consumption. Giacomo Bescemi et al. \cite{BUSCEMI20251349} used Deep reinforcement learning-based control of thermal energy storage for university classrooms and demonstrated that DRL apprach achieved a 3.2\% decrease in operating cost compared to rule based controls.

In this work we consider an HVAC cooling plant consisting of:
(i) a fixed-capacity chiller with controllable part-load ratio (PLR), and
(ii) a TES system that stores ``units of refrigeration'' (cooling energy) and later discharges to meet building cooling demand.
At each hour, the operator must meet an exogenous cooling load requirement. The operator chooses the chiller PLR, which determines chiller cooling output for that hour; the mismatch between chiller output and load is absorbed by TES charge/discharge subject to efficiency and capacity constraints.

The design objective is not only operational cost minimization for a fixed plant, but also \emph{sizing}: choosing the chiller capacity and TES capacity to minimize the \emph{30-year life-cycle cost} (LCC). A key economic characteristic in our setting is that increasing chiller capacity is significantly more capital-intensive than increasing TES capacity: a unit increase in chiller capacity incurs approximately 4.3 times the capital cost of the same unit increase in TES capacity. \cite{LUERSSEN2019640} Therefore, identifying the best combination of chiller and TES capacity under an optimal operating policy is non-trivial.

\textbf{Approach.} We treat hourly plant operation as a Markov Decision Process (MDP), solved using a Deep Q Network (DQN)\cite{dqn-original-paper}. For each candidate sizing configuration, we train a DQN policy on historical traces of cooling load and electricity prices and evaluate (i) annual electricity cost and (ii) loss-of-load (unmet cooling) metrics. We then compute the 30-year LCC and select the cost-minimizing sizing configuration among those that satisfy the cooling load without loss-of-load.

\textbf{Contributions.} This paper makes the following contributions:
\begin{itemize}
  \item We formulate a constrained MDP for chiller PLR control with TES dynamics and feasibility constraints.
  \item We implement a DQN with a maskable action space to enforce operational feasibility.
  \item We evaluate across multiple chiller/TES sizing configurations and quantify annual electricity cost and loss-of-load tradeoffs (Table~\ref{tab:performance-best}).
  \item We couple learned operation with a 30-year discounted life-cycle cost model to select a cost-minimizing sizing configuration among feasible candidates.
\end{itemize}


\section{System Model and Life-Cycle Cost}
\subsection{Cooling Infrastructure}

The cooling infrastructure consists of two components:
\paragraph{Chiller:} The chiller is a fixed-capacity electric unit with rated thermal output
$C_{\text{ch}}$ (kWh$_{\text{th}}$). At each hour $k$, the chiller is operated at a part-load ratio (PLR)
$a_k \in [0,1]$, producing $a_k C_{\text{ch}}$ units of cooling energy. The electrical power consumed by the chiller depends on the PLR and ambient temperature through manufacturer-specified efficiency curves.
\paragraph{Thermal Energy Storage (TES):} The TES stores cooling energy (e.g., chilled water or phase-change material) and discharges it to meet demand. Let $E_k$ denote the state of charge (SoC) of the TES at hour $k$, measured in kWh$_{\text{th}}$. The SoC is bounded as
\[
0 \le E_k \le E_{\max},
\]
where $E_{\max}$ is the TES capacity.

At each hour, the HVAC cooling demand must be met by the combined output of the chiller and the TES discharge.
The goal is to minimize the life-cycle cost of the HVAC system.  This life-cycle cost is made up of both capital expenses (or investments) made towards acquiring the HVAC infrastructure and operational costs, i.e., electricity for operating chillers and maintenance. The HVAC system consists of two parts, a chiller and thermal energy storage. The chiller generates cold water at 4$^o$C, while the thermal energy storage stores this water. These two together satisfy cooling load demands, which vary based on hour of day as well as time of year. 

Naturally, the larger the size of the chiller and thermal energy storage, the larger the up-front capital expenditure. However, with the combination of both a chiller and thermal energy, this allows for predictive scheduling of thermal energy storage when electricity costs are low, as well as in anticipation for a cooling load spike. What this intelligent schedule allows is a chiller capacity which is lower than the peak cooling demand. The overall goal of this endeavour is to find a combination of chiller capacity and thermal electric storage which minimizes this overall life-cycle cost while (almost) always maintaining the cooling load.

\subsection{Life-Cycle Cost Model}

We consider a life-cycle horizon of 30 years. The total life-cycle cost (LCC) is
\begin{equation}\label{eq:lcc}
LCC
=
CAPEX_{\text{ch}}
+
CAPEX_{\text{TES}}
+
OPEX,
\end{equation}
where capital expenditures ($CAPEX_{\text{ch}}$ and $CAPEX_{\text{TES}}$) are proportional to installed capacities, and operating expenditure $OPEX$ includes electricity and maintenance costs. The two capital expenditures are just the product of the cost per $kW_{th}$ and the chosen chiller and TES storage capacities. The operational expenditure takes a more complicated form.

Let $f_{\text{elec}}(C_{\text{ch}}, E_{\max})$ denote the annual electricity cost incurred when operating a plant with chiller capacity $C_{\text{ch}}$ and TES capacity $E_{\max}$ under an optimal policy. Maintenance costs are modeled as a fixed fraction of the total capital expenditure. That is set as 2\% of the total CAPEX, which is subject to an inflation of 5\% year-over-year. The net present value of operating expenditure is
\begin{equation}\label{eq:opex}
OPEX
=
\sum_{i=1}^{30}
\frac{
f_{\text{elec}}
+
0.02\, CAPEX_{\text{total}} \, 1.05^{\,i-1}
}{
1.06^{\,i-1}
},
\end{equation}
where maintenance costs are assumed to be $2\%$ of the total capital expenditure $CAPEX_{\text{total}}$ and grow at an annual inflation rate of $5\%$. We compute the net present value of operating expenditure using a $6\%$ discount rate \cite{LUERSSEN2019640}.

The sizing problem is to find the pair $(C_{\text{ch}}, E_{\max})$ that minimizes the life-cycle cost $LCC$.

\section*{Problem Formulation}
\subsection{Markov Decision Process (MDP) Model}
 We pose this as a Markov Decision Process (MDP) problem for the chiller part-load-ratio (PLR) control with thermal energy storage dynamics and feasibility constraints.
For fixed $(C_{\text{ch}}, E_{\max})$, we model hourly operation as a discrete-time Markov decision process (MDP). The MDP model is a tuple $\{\mathcal{S},\mathcal{A}, \mathcal{R}, \mathcal{T}, \gamma \},$ where $\mathcal{S}$ is the state space, $\mathcal{S}$ is action space, $\mathcal{S}$ is transition probability matrix and $\gamma \ (0,1)$ is the MDP discount factor.

\paragraph{State Space}
 At hour $k$, the system state is
\begin{equation}
S_k = (X_k, E_k, P_k, h_k, d_k, \mathbf{A}_k),
\end{equation}
where $X_k$ denotes the cooling demand, $E_k$ is the state of charge (SoC) of the TES, $P_k$ is the electricity price ($P_k$ is the least price among the following $M$ electricity sources), $h_k$ is the hour of the day, $d_k$ is the day index, and $\mathbf{A}_k=[\text{ElecSrc}_1,\text{ElecSrc}_2,\dots, \text{ElecSrc}_M], \text{ElecSrc}_i \in \{0,1\},$  is a binary variable vector describing which of the $M$ electricity sources are available at time $k$. Cooling demand and electricity prices evolve exogenously based on historical data.



\paragraph{Action Space and Feasibility Constraints}

At each time step $k$, the control action is the chiller part-load ratio (PLR),
which determines the instantaneous cooling/ refrigeration production $a_k C_{\text{ch}}$, where
$C_{\text{ch}}$ is the rated chiller capacity. Not all action choices (value of PLR) are feasible at every time step. Feasibility is determined jointly by the current cooling demand $X_k$, the TES state of charge $E_k$, the TES capacity $E_{\max}$, and the charging--discharging efficiency $\eta \in (0,1]$. Specifically, the chiller output together with the maximum admissible TES discharge must be sufficient to meet the cooling demand, while any excess cooling production must not exceed the available TES charging capacity and the available headroom to store in it.

The feasible action set at time $k$ is therefore given by
\begin{equation} 
\mathcal{A}_{\text{feas}}(S_k)
=
\left\{
a \in \mathcal{A} :
\underline{a}_k \le a \le \overline{a}_k
\right\},
\label{eq:feasible_actions}
\end{equation}
where the lower and upper bounds are defined as
\begin{align}
\underline{a}_k
&=
\max\!\left(
0,\;
\frac{X_k - \eta E_k}{C_{\text{ch}}}
\right), \label{eqn:action-constraint2}
\\
\overline{a}_k
&=
\min\!\left(
1,\;
\frac{X_k + (E_{\max} - E_k)/\eta}{C_{\text{ch}}}
\right). \label{eqn:action-constraint3}
\end{align}

The lower bound $\underline{a}_k$ ensures that, even after fully discharging the TES (accounting for efficiency losses), the combined cooling output is sufficient to meet the load $X_k$. The upper bound $\overline{a}_k$ ensures that any excess cooling production can be absorbed by the TES without exceeding its maximum capacity.

These constraints guarantee that the TES state of charge remains within the admissible interval $[0, E_{\max}]$ at all times.

\paragraph{How this connects to loss-of-load}
A loss-of-cooling-load event occurs when $\underline{a}_k > \overline{a}_k,$ i.e., when the cooling demand cannot be met even if the chiller operates at full capacity and the TES is fully discharged. This situation arises when the chiller is undersized relative to peak demand and the TES is insufficiently charged\footnote{loss of load can also occur due to a sub-optimal policy that completely ignores it and still satisfies feasibility constraints}. During such event, we operate the chiller at max capacity (PLR=1 or $a_k=1$).

This condition is precisely what motivates: the loss-of-load penalty in the reward function (for action $a_k=1$), and the restriction to zero-loss-of-load configurations in the sizing optimization.

\subsection*{Loss-of-Cooling-Load During Operation and Penalty Design}
\paragraph{Design of Reward Function and Loss-of-Load Modeling}

For a fixed infrastructure sizing $(C_{\text{ch}}, E_{\max})$, the capital and maintenance costs are constants and therefore do not affect short-term operational decisions. The operational objective of the Markov decision process is thus to minimize the cumulative electricity expenditure over the evaluation horizon.

Accordingly, the baseline instantaneous reward associated with the state--action pair $(S_k, a_k)$ is defined as the negative electricity cost incurred during hour $k$:
\begin{equation}
R_k^{\text{elec}}
=
- P_k \cdot \text{ElecPower}(a_k),
\end{equation}
where $P_k$ denotes the electricity price and $\text{ElecPower}(a_k)$ is the electrical power consumed by the chiller when operated at part-load ratio $a_k$.

For certain sizing configurations, particularly when the chiller capacity is smaller than the peak cooling demand, there may exist states $S_k$ for which no feasible control action can fully satisfy the cooling load. This occurs when the maximum available cooling from the chiller and the TES is insufficient to meet demand. Formally, a loss-of-cooling-load event occurs whenever
$
X_k > a_k C_{\text{ch}} + \eta E_k,
$

We define the instantaneous loss-of-load magnitude as
\begin{equation}
\ell_k \triangleq \max\!\left(0,\; X_k - a_k C_{\text{ch}} - \eta E_k \right),
\end{equation}
which represents the unmet portion of the cooling demand during hour $k$.

During policy evaluation, loss-of-load is recorded as a reliability metric but does not directly affect the realized electricity cost. During training, however, loss-of-load must be explicitly penalized in order to discourage infeasible operation. The training reward is therefore augmented as
\begin{equation} \label{eq:reward-with-penalty}
R_k
=
- P_k \cdot \text{ElecPower}(a_k)
- \lambda \, \ell_k,
\end{equation}
where $\lambda > 0$ is a penalty coefficient chosen sufficiently large to ensure that incurring loss-of-load is always more costly than preemptively discharging the TES or operating the chiller at a lower PLR in earlier periods. Note that this penalty should be significantly greater than the potential cost of anticipating price variation and charging the thermal energy storage. An RL agent would not avoid this loss-of-load if the incurred penalty
is not greater than the cost of charging the energy storage.

This reward structure induces policies that minimize electricity cost while avoiding loss-of-cooling-load whenever feasible. In the outer-loop sizing optimization, only configurations that result in zero cumulative loss-of-load over the evaluation horizon are deemed feasible and considered for life-cycle cost minimization. (see resource optimization Section~\ref{sec:resource_sizing})
 
Note that for the chosen action $a_k$, i.e., part-load-ratio $(PLR)$ of the chiller operation, the equivalent electric power consumption $\text{ElecPower}(a_k)$ in $KW$ is a nonlinear function of $a_k$ given by
\begin{equation}
        \text{ElecPower}(a_k)=    
        Q_{ref} \times EIRPLR(a_k) \times CAPFT \times ERIFT
\end{equation}
where $EIRPLR$ is `electric input to cooling output ratio function' of part-load-ratio curve, $CAPFT$ is cooling capacity function of temperature curve, $EIRFT$ is `electric input to cooling output ratio function' of temperature curve, $COP(a_k)$ is the coefficient of performance defined as the ratio of thermal cooling delivered at the output of the chiller by the input electric power applied at part-load-ratio $a_k$ and,  $Q_{ref}$ is the maximum electric power consumption of the chiller at full loading \cite{energyplus2025}. These values  can be inferred from the chiller datasheet supplied by the manufacturer.


\paragraph{State transition: TES Dynamics}
The TES state evolves deterministically according to chosen action $a_k$ as
\begin{equation}
E_{k+1}
=
E_k
+
\begin{cases}
\eta \bigl(a_k C_{\text{ch}} - X_k\bigr),
& a_k C_{\text{ch}} \ge X_k, \\[6pt]
\bigl(a_k C_{\text{ch}} - X_k\bigr)/\eta,
& a_k C_{\text{ch}} < X_k,
\end{cases}
\end{equation}
where $\eta$ denotes the charging/discharging efficiency. We assume a round-trip efficiency of $0.9$ with equal charge and discharge efficiencies, which implies $\eta=\sqrt{0.9}\approx0.949$. The constraint on action imposed in \eqref{eq:feasible_actions}--\eqref{eqn:action-constraint3}  ensures that the dynamics of the energy storage level described above do not go below zero level or above the maximum TES capacity, while also taking the round-trip efficiency into consideration. 

Other components of the state variable $S_k$ evolves stochastically as the exhogenous input to the system.
The hourly building cooling load $X_k$, electricity prices $P_k$, and availability of electricity sources $\mathbf{A}_k$ evolve in a stochastic manner as a discrete-time Markov chain over a finite state space. 

\subsection{Why RL is needed:}

\paragraph{Difficulties in solving this MDP problem formulated above}
The MDP setting formulated above to find an optimal policy of operation is extremely hard. The action space is constrained by the current state, which makes both finding an optimal policy to carry out the operation as well as learning it via RL training a harder problems to solve.
Since the action space is explicitly state-dependent (due to TES feasibility constraints), it violates the assumptions underlying classical dynamic programming and significantly complicates analytical policy derivation. Reinforcement learning provides a natural framework for learning cost-to-go functions under such constrained, non-stationary dynamics directly from data.  The details about the nontriviality of optimal policy and justification for use of RL are given in Section~\ref{sec:policy_hierarchy}.

\section{Chiller Operation Methodologies and Policy Structure}\label{sec:all_policies}
This section develops a structured understanding of operational control for the chiller–TES system model introduced in the previous Section. Our goal is twofold. First, we design (and then define concretely) a set of baseline control policies that span a wide range of operational behaviors, from purely reactive to highly conservative. Second, we use these policies to expose the structural difficulty of computing an optimal operating policy of the MDP setting, thereby motivating the reinforcement learning approach adopted in Section~\ref{sec:RL method}.

Throughout this section, we consider a fixed infrastructure configuration 
$(C_{\text{ch}}, E_{\text{max}})$ and focus exclusively on the operational decision problem.
\subsection*{Operational Tradeoff}

At each hour $k$, the system operator must select a chiller part-load ratio $a_k$ that determines the cooling production $a_k C_{\mathrm{ch}}$. Once the choice of $a_k$ is done by the adopted scheduling policy, any mismatch between production and the building cooling demand $X_k$ is absorbed by charging or discharging the thermal energy storage (TES), subject to efficiency and capacity constraints.

A key feature of this problem is that TES couples decisions across time. Using TES to reduce electricity cost in the present hour may reduce feasibility (incur loss of load later) or increase cost in future hours, while charging TES incurs present cost in exchange for future flexibility. This intertemporal coupling is further complicated by  time-varying electricity prices, nonstationary cooling demand (since it varies over season), nonlinear electricity consumption of the chiller as a function of part-load ratio and temperature, and state-dependent feasibility constraints~\eqref{eq:feasible_actions}--\eqref{eqn:action-constraint3}. As a result, the operational problem is inherently non-myopic.

To build intuition, we first introduce three baseline policies, namely, \emph{Greedy}, \emph{TES-First Feasibility}, and \emph{Storage-Dominant Pessimistic}, that reflect distinct assumptions about the value of stored cooling energy. We then contrast these with the optimal policy.

\subsection{Greedy (Myopic) Policy} \label{subsec:greedy-description}

The \emph{Greedy (Myopic) Policy} minimizes instantaneous electricity cost without regard to future consequences, trying only to meet the present cooling demand. Formally, at time $k$, the greedy policy selects
\begin{equation}
a_k^{\mathrm{greedy}}
=
\arg\min_{a \in \mathcal{A}_{\mathrm{feas}}(S_k)}
P_k \cdot \mathrm{ElecPower}(a),
\end{equation}
where $\mathcal{A}_{\mathrm{feas}}(S_k)$ is the feasible action set defined in~\eqref{eq:feasible_actions}--\eqref{eqn:action-constraint3}.

Because electricity prices are nonnegative and chiller power consumption is nondecreasing in the part-load ratio (PLR), the greedy policy always selects the minimum feasible PLR,
\begin{equation}
a_k^{\mathrm{greedy}}
=
\max\!\left(0,\,
\frac{X_k - \eta E_k}{C_{\mathrm{ch}}}
\right),
\end{equation}
which is exactly the lower bound $\underline{a}_k$ in \eqref{eqn:action-constraint2}. In other words, the greedy policy always acts by taking the lower bound $\underline{a}_k$ of feasible actions $\mathcal{A}_{\text{feas}}(S_k)
=
\left\{
a \in \mathcal{A} :
\underline{a}_k \le a \le \overline{a}_k
\right\}.$ Under the greedy policy TES is discharged whenever possible, the chiller is used only when storage is insufficient, and TES is never charged unless forced by feasibility.

This policy implicitly treats stored cooling energy as having zero future value (and hence discards any opportunity to store it). While it often yields low electricity cost in the short term, it performs poorly during demand spikes or prolonged high-price periods, frequently incurring loss-of-cooling-load unless the chiller is sized close to peak demand.

While the myopic policy generates a low electricity cost, it is observed to have a fairly large loss-of-load. The policy does not predict spikes in cooling load, and consequently charge the TES. As a result, this policy would require a chiller capacity of around the range of the peak cooling load to prevent a loss of cooling load.

At each hour, the myopic policy selects the chiller part-load ratio that minimizes instantaneous electricity cost subject to feasibility constraints. Equivalently, it chooses the smallest feasible part-load ratio that satisfies the current cooling demand given the available TES state of charge. This policy does not anticipate future demand or electricity prices and therefore does not charge thermal energy storage except when forced by feasibility.

\subsection{TES-First Feasibility Policy (TFP)}
\begin{figure}
\includegraphics[width=0.65\linewidth, trim=10pt 0pt 15pt 0pt, clip]{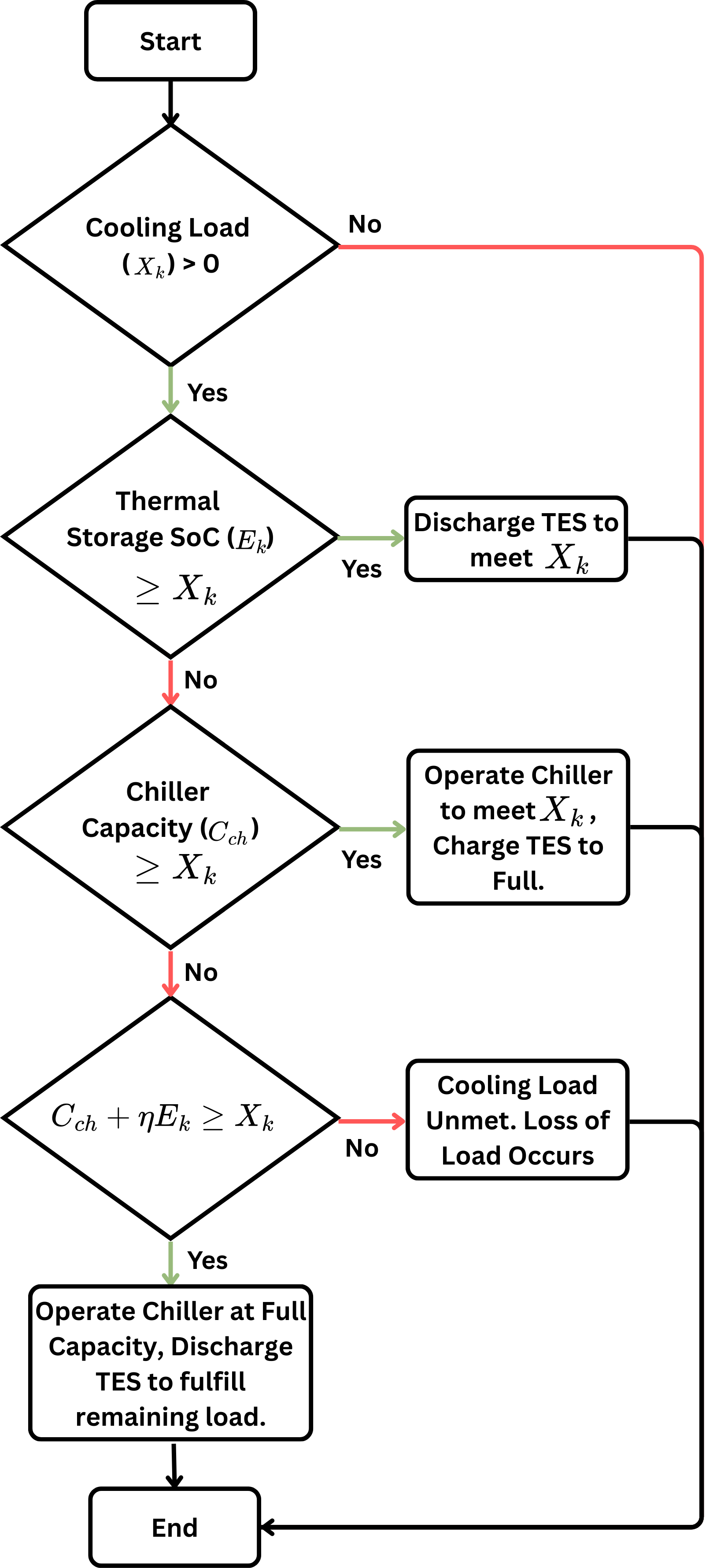}
  \caption{Control logic for TES-First Feasibility Policy (TFP)}
  \label{fig:decisiontree}
\end{figure}

We designed a new policy called the TES-First Feasibility Policy (TFP) that heuristically tries to optimize the operation. As mentioned, finding the true optimal MDP policy is prohibitively hard in our problem. We find that this TFP policy matches very closely to the performance of the RL policy, which aims to learn the optimal policy (see experimental evaluation Section~\ref{sec:simulations}).

At each hour, TFP operates according to the following rules:
\begin{enumerate}
    \item If TES alone can meet the demand ($E_k \ge X_k$), discharge TES fully and do not operate the chiller.
    \item Otherwise, if the chiller alone can meet the demand ($C_{\mathrm{ch}} \ge X_k$), operate the chiller to meet demand and use any remaining capacity to charge TES to the maximum possible extent.
    \item If both chiller and TES are required, operate the chiller at full capacity and discharge TES to meet the remaining demand.
    \item If the combined capacity of chiller and TES is insufficient, record an unmet-load event.
\end{enumerate}

This policy is illustrated by the decision logic in Fig.~\ref{fig:decisiontree}. The TFP policy serves as a conservative baseline that prioritizes the reliable satisfaction of cooling demand.  
Under TFP, storage is used immediately when available ($E_k \ge X_k$) to the maximum possible extent and, therefore, is not preserved for future benefit. However, if $E_k<X_k,$ then storage is immediately charged to the maximum possible extent by operating the chiller at the maximum possible part-load ratio, thereby preserving it for future benefit (see Fig.~\ref{fig:decisiontree}).  As a result, it provides a conservative and interpretable baseline that attempts to avoid loss-of-load, but may incur high electricity cost by failing to exploit temporal variations in electricity prices. 
However, prioritizing the use of the TES to satisfy cooling demands lacks foresightedness about consistent high cooling loads that may occur in the near future, 
possibly leading to loss-of-load scenarios. 

We note that the TFP policy selects action as
\begin{equation}
    a^{TFP}_k=\begin{cases}
        0 \quad E_k \ge X_k,\\
        \overline{a}_k, \quad E_k <X_k
    \end{cases},
\end{equation}
where $\overline{a}_k$ is given by \eqref{eqn:action-constraint3} and is the upper bound of feasible actions set $\mathcal{A}_{\text{feas}}(S_k).$

\subsection{Storage-Dominant Pessimistic Policy (SDPP): A Theoretical Extreme}
To characterize the structure of optimal control, it is useful to consider a theoretical pessimistic extreme that tries to preserve the stored energy at all possible times. The \emph{Storage-Dominant Pessimistic Policy (SDPP)} is defined as the policy that, among all feasible actions, selects one that maximizes the next-step TES state of charge:
\begin{equation}\label{eq:SDPP}
a_k^{\mathrm{SDPP}}
=
\arg\max_{a \in \mathcal{A}_{\mathrm{feas}}(S_k)}
E_{k+1}(a),
\end{equation}
subject to meeting cooling load. It never voluntarily discharges TES when an alternative feasible action exists that preserves or increases storage, regardless of electricity price. If we compare \eqref{eq:SDPP} with \eqref{eqn:action-constraint2}, it is clear that the SDPP policy always acts by taking the upper bound $\overline{a}_k$ of feasible actions $\mathcal{A}_{\text{feas}}(S_k)
=
\left\{
a \in \mathcal{A} :
\underline{a}_k \le a \le \overline{a}_k
\right\}$ in order to operate the chiller at maximum possible part-load-ratio ($a_k$) without wasting any cooling energy generated at chiller output. 

We emphasize that SDPP serves as a theoretical extreme that bounds policy behavior and clarifies the structure of the optimal policy which we describe next.

\subsection{Optimal Policy and Marginal Value of Storage} \label{subsec:optiaml-policy-and-marginal-value}

The optimal policy $\pi^\star$ minimizes cumulative discounted electricity cost subject to feasibility constraints and is characterized by the Bellman equation (with value function $V(s)$ representing cost)
\begin{equation}
V_k(S_k)
=
\min_{a \in \mathcal{A}_{\mathrm{feas}}(S_k)}
\Bigl[
P_k \cdot \mathrm{ElecPower}(a)
+
\mathbb{E}\!\left[ V_{k+1}(S_{k+1}) \right]
\Bigr].
\end{equation}
Here the next state evolve as
$
S_{k+1}
=
g\!\left( S_k, a_k, W_{k+1} \right),
$
where $g(\cdot)$ represents the system dynamics, including the TES state update with time, and $W_{k+1}$ denotes exogenous information such as the next-period cooling demand and electricity price.

We define a useful quantity \emph{marginal value of stored cooling energy} for understanding optimal behavior as :
\begin{equation}
\lambda_k(E_k)
\;\triangleq\;
V_k(E_k - 1, \cdot) - V_k(E_k, \cdot).
\end{equation}
This quantity represents the expected future electricity cost avoided by retaining one additional unit of TES energy at time $k$. Under this definition of marginal value of storage, we have
\begin{enumerate}
    \item the greedy policy corresponds to $\lambda_k = 0$;
    \item TFP corresponds to a finite but non-anticipatory $\lambda_k$;
    \item SDPP corresponds to $\lambda_k \to \infty$; and
    \item the optimal policy assigns a finite, time- and state-dependent $\lambda_k$.
\end{enumerate}

\subsection{Policy Hierarchy and Structural Result}\label{sec:policy_hierarchy}

The four policies introduced above admit a natural ordering:
\begin{equation}
\lambda^{\mathrm{greedy}}
<
\lambda^{\mathrm{TFP}}
<
\lambda^\star
<
\lambda^{\mathrm{SDPP}}.
\end{equation}

This ordering reflects increasing valuation of stored cooling energy and increasing anticipatory behavior. The greedy policy minimizes immediate cost, TFP prioritizes feasibility with some anticipation, SDPP maximizes storage preservation regardless of cost, and the optimal policy lies strictly between these extremes.

Although the TES-First Feasibility Policy (TFP) assigns a positive value to stored cooling energy by charging TES whenever immediate demand cannot be met, it does so without anticipating future electricity prices or cooling demand. In particular, TFP discharges TES whenever $E_k \ge X_k,$
regardless of whether future periods exhibit higher prices or tighter feasibility constraints, and charges TES whenever $ E_k < X_k,$ regardless of whether future conditions suggests preservation of storage. As a result, TFP assigns a finite but non-anticipatory marginal value to storage, reflecting feasibility rather than future cost avoidance. Since the optimal policy explicitly trades off current electricity expenditure against the expected future value of stored energy, it follows that $\lambda^{\mathrm{TFP}} < \lambda^\star .$

\subsection*{Why Finding the Optimal Policy Is Nontrivial}

Despite this clear structure, computing the optimal policy analytically is intractable. The difficulties arise from state-dependent action feasibility,  nonlinear and temperature-dependent chiller efficiency (chiller output vs electricity consumed), nonstationary demand and price processes since they both vary across seasons; and the high dimensionality of the state space. Classical dynamic programming assumptions do not apply, and closed-form threshold policies are unavailable. At the same time, purely reactive heuristics (greedy, TFP) fail to balance electricity cost and future feasibility.

This motivates the use of reinforcement learning as a numerical method for approximating the optimal cost-to-go function and the associated implicit valuation of thermal energy storage.

\subsection*{Role of Reinforcement Learning}

Reinforcement learning bridges the gap between heuristic baselines and the optimal policy by learning  (from data) how much stored cooling energy should be valued in different states and at different times. The DQN policy introduced in the next section approximates this valuation without requiring explicit modeling of future demand or price distributions. In this sense, RL is used as a principled tool for estimating the minimum achievable operational cost under optimal anticipatory control. This estimate is essential for the infrastructure sizing optimization described in Section~\ref{sec:resource_sizing}.

A trained RL agent (or policy), unlike the previous policies, does (implicitly) track the variation in the cooling load as well as price. As a result, it can intelligently schedule the chiller operation to charge the thermal energy storage when it anticipates a spike in cooling load, as well as charge the thermal energy storage when electricity prices are low and discharge the thermal energy storage when the electricity prices are high.

\section{Reinforcement Learning Framework}\label{sec:RL method}

\begin{figure}
    \centering
    \includegraphics[width=0.78\linewidth]{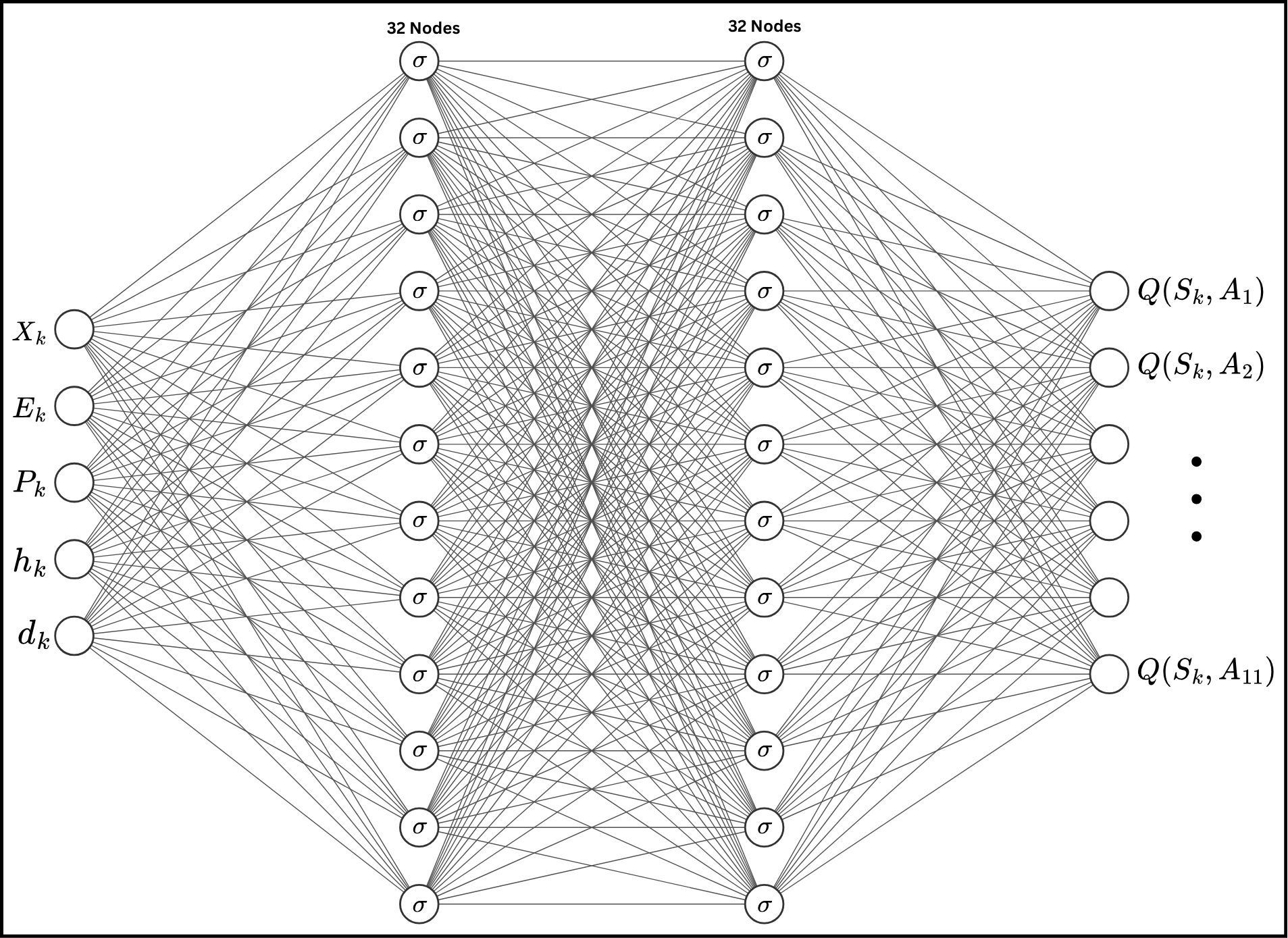}
    \caption{Architecture diagram of DQN}
    \label{fig:dqn-diagram}
\end{figure}

This section describes the reinforcement learning framework used to compute an operational policy for the chiller--TES system introduced in Section~\ref{sec:all_policies}. We first provide a brief overview of the vanilla Deep Q-Network (DQN) algorithm. Implementation details and training hyperparameters are deferred to Section~\ref{sec:simulations}.

\subsection{Background: Deep Q-Networks}

Deep Q-Networks (DQN) are a class of value-based reinforcement learning methods designed to approximate the optimal action--value function for large or continuous state spaces. Given an MDP with state $S_k$, action $a_k$, reward $R_k$, and transition dynamics, the optimal action--value function is defined as
\begin{equation*}
Q_k^\star(S_k,a_k)
=
\mathbb{E}\!\left[
\sum_{t=k}^{T} R_t
\,\middle|\,
S_k,a_k,\pi^\star
\right],
\end{equation*}
where $\pi^\star$ is an optimal policy.

In DQN, the function $Q_k^\star(\cdot,\cdot)$ is approximated using a neural network parameterized by weights $w$, denoted $Q(S_k,a_k;w)$ (see Fig. \ref{fig:dqn-diagram}). The network is trained to satisfy the Bellman optimality condition:
\begin{equation*}
Q^\star(S_k,a_k)
=
R(S_k,a_k)
+
\mathbb{E}\!\left[
\min_{a' \in \mathcal{A}(S_{k+1})}
Q^\star(S_{k+1},a')
\right].
\end{equation*}

To stabilize training, DQN employs (i) experience replay, which breaks correlations between consecutive samples, and (ii) a target network that provides a slowly varying estimate of the Bellman target.

Because the feasible action set depends on the current state, the Bellman update must explicitly account for feasibility. The optimal value function satisfies
\begin{equation*}
V_k(S_k)
=
\min_{a \in \mathcal{A}_{\mathrm{feas}}(S_k)}
\left[
P_k \cdot \mathrm{ElecPower}(a)
+
\mathbb{E}\!\left[
V_{k+1}(S_{k+1})
\right]
\right],
\end{equation*}
with state evolution
$
S_{k+1}
=
f(S_k,a_k,W_{k+1}),
$
 where $f(\cdot)$ represents the TES dynamics and time advancement, and $W_{k+1}$ denotes exogenous demand and price realizations.

In the DQN implementation, infeasible actions are excluded by masking their Q-values before action selection and Bellman minimization. This ensures that learning is performed entirely within the physically admissible region of the action space.

\subsection{Training Procedure}

Training proceeds over multiple episodes, where each episode corresponds to a full year of hourly operation. Within an episode, the agent observes the current state $S_k$, selects an action $a_k$ using an exploration policy, observes the resulting reward and next state, and stores the transition in an experience replay buffer.

Mini-batches of transitions are sampled from the replay buffer to update the Q-network by minimizing the temporal-difference error between predicted Q-values and Bellman targets computed using a separate target network. Exploration is gradually reduced over training to encourage convergence to a deterministic policy. 

The same historical traces of cooling demand and electricity prices are used across episodes, while the initial TES state is randomized to improve coverage of the state space. Since the learned policy is used for offline evaluation and design optimization rather than real-time deployment, overfitting to historical traces is acceptable in this context.

The trained RL policy pick the action as 
\[
\pi_{\text{RL}}(s) = \arg\max_{a \in \mathcal{A}(s)} \hat{Q}(s,a;w),
\]
where ${Q}(s,a;w)$ is the post training $Q$-function learned through Algorithm 1. 
The specific hyperparameter values and training configurations used in
the experiments are reported in Section~\ref{sec:simulations}. 

\begin{algorithm}
\caption{Vanilla DQN for Chiller--TES Operation}
\label{alg:dqn}
\begin{algorithmic}[1]
\Require
Resource capacity $\theta = (C_{\mathrm{ch}}, E_{\max})$;
feasible action bounds $\underline a_k, \overline a_k$ from (5)--(6);
historical demand and price traces $\{X_k, P_k\}_{k=1}^T$
\Ensure
Trained Q-network parameters $w$

\State Initialize Q-network $Q(S,a;w)$ with random weights
\State Initialize target network $\bar Q(S,a;\bar w) \gets Q(S,a;w)$
\State Initialize replay buffer $\mathcal{D}$

\For{each training episode}
    \State Sample initial TES state $E_1$
    \State Initialize state $S_1 = (X_1, E_1, P_1, h_1, d_1)$
    \For{$k = 1$ to $T$}
        \State Compute feasible action set $\mathcal{A}_{\mathrm{feas}}(S_k) = [\underline a_k, \overline a_k]$
        \State Select action $a_k$ using $\epsilon$-greedy policy over $\mathcal{A}_{\mathrm{feas}}(S_k)$:
        \[
        a_k =
        \begin{cases}
        \text{random action in } \mathcal{A}_{\mathrm{feas}}(S_k), & \text{with prob. } \epsilon \\
        \arg\min_{a \in \mathcal{A}_{\mathrm{feas}}(S_k)} Q(S_k,a;w), & \text{with prob. } 1-\epsilon
        \end{cases}
        \]
        \State Compute loss-of-load
\[
\ell_k = \max\!\left\{0,\; X_k - \big(a_k C_{\mathrm{ch}} + E_k^{\mathrm{dis}}\big) \right\}
\]

        \State Observe reward $R_k = - P_k \cdot \mathrm{ElecPower}(a_k) - \lambda \, \ell_k$
        \State Observe next state $S_{k+1} = f(S_k, a_k, W_{k+1})$
        \State Store transition $(S_k, a_k, R_k, S_{k+1})$ in $\mathcal{D}$
        \State Sample mini-batch $\{(S_i,a_i,R_i,S'_i)\}$ from $\mathcal{D}$
        \State Compute target values:
        \[
        y_i = R_i + \min_{a' \in \mathcal{A}_{\mathrm{feas}}(S'_i)} \bar Q(S'_i,a';\bar w)
        \]
        \State Update $w$ by minimizing:
        \[
        \sum_i \left( Q(S_i,a_i;w) - y_i \right)^2
        \]
        \State Periodically update target network $\bar w \gets w$
    \EndFor
\EndFor
\end{algorithmic}
\end{algorithm}

\subsection{Role of Reinforcement Learning in the Co-Design Problem}

The reinforcement learning framework described above is not used merely to control the chiller, but as a computational tool to evaluate the operational cost of a given infrastructure configuration. For each candidate pair $(C_{\mathrm{ch}},E_{\max})$, a DQN is trained to approximate the optimal operational policy and associated electricity cost.

These learned costs are then combined with capital expenditure in the outer-loop sizing optimization described in Section~\ref{sec:resource_sizing}. In this sense, reinforcement learning enables the estimation of the optimal policy which cannot be obtained analytically due to nonstationary demand, nonlinear electricity consumption, and state-dependent feasibility constraints.


\section{Optimal Sizing of Cooling Infrastructure}\label{sec:resource_sizing}


The problem of determining the cost-optimal cooling infrastructure is decomposed into two coupled but conceptually distinct layers:
\begin{itemize}
\item \textbf{Operational layer (inner loop):}
For a fixed infrastructure sizing $(C_{\text{ch}}, E_{\max})$, determine an operating policy that minimizes electricity cost subject to physical feasibility and loss-of-cooling-load constraints.

\item \textbf{Sizing layer (outer loop):}
Select the chiller and TES capacities that minimize total life-cycle cost, assuming optimal operation at the inner layer.
\end{itemize}

This separation avoids mixing long-term capital investment decisions with short-term operational control. This also enables reinforcement learning to be used as a cost-evaluation tool rather than as a direct sizing optimizer.

\subsection{Feasible Sizing Configurations}

Let $\mathcal{C}_{\text{ch}}$ and $\mathcal{E}_{\max}$ denote discrete candidate sets of chiller and TES capacities, respectively. For each candidate sizing configuration
\begin{equation}
\theta = (C_{\text{ch}}, E_{\max}) \in \mathcal{C}_{\text{ch}} \times \mathcal{E}_{\max},
\end{equation}
we train a DQN-based operational policy $\pi_{\theta}$ as described in Section~\ref{sec:RL method}. The trained policy $\pi_{\theta}$ is evaluated over the full one-year historical trace of cooling demand and electricity prices. Let $\ell_k^{\theta}$ denote the instantaneous loss-of-cooling-load incurred at hour $k$ under policy $\pi_{\theta}$. The cumulative loss-of-load over the evaluation horizon $T$ is
\begin{equation}
L(\theta)
=
\sum_{k=1}^{T} \ell_k^{\theta},
\qquad T = 8760.
\end{equation}

We define the set of feasible sizing configurations as
\begin{equation}
\Theta_{\text{feas}}
=
\left\{
\theta : L(\theta) = 0
\right\}.
\end{equation}
That is, only those infrastructure configurations for which the learned policy meets the cooling demand at all time steps are considered admissible. This restriction enforces cooling reliability as a hard constraint rather than a soft trade-off against cost.

\subsection{Estimation of Operational Cost}

For each feasible sizing configuration $\theta \in \Theta_{\text{feas}}$, the trained policy $\pi_{\theta}$ yields an estimate of the annual electricity cost
\begin{equation}
f_{\text{elec}}(\theta)
=
\sum_{k=1}^{T}
P_k \cdot \text{ElecPower}\!\left(a_k^{\pi_{\theta}}\right),
\end{equation}
where $a_k^{\pi_{\theta}}$ denotes the action selected by policy $\pi_{\theta}$ at time $k$. With the accurately trained RL agent, this cost estimate represents the minimum achievable electricity expenditure for configuration $\theta$ under the observed realization of cooling demand and electricity prices.

\subsection{Life-Cycle Cost Evaluation}
Recall that, from \eqref{eq:lcc} and \eqref{eq:opex}, the total life-cycle cost associated with configuration $\theta= (C_{\text{ch}}, E_{\max})$ is given by
\begin{equation}
LCC(\theta)
=
CAPEX_{\text{ch}}(C_{\text{ch}})
+
CAPEX_{\text{TES}}(E_{\max})
+
OPEX(\theta),
\end{equation}
where the operating expenditure $OPEX(\theta)$ is computed as the discounted net present value of electricity and maintenance costs over a 30-year horizon as
\begin{equation}
OPEX(\theta)
=
\sum_{i=1}^{30}
\frac{
f_{\text{elec}}(\theta)
+
0.02 \cdot CAPEX_{\text{total}}(\theta) \cdot 1.05^{\,i-1}
}{
1.06^{\,i}
}.
\end{equation}

This formulation explicitly captures the trade-off between higher upfront investment and reduced long-term operating cost enabled by operational flexibility.

\subsection{Optimal Sizing Selection}

The optimal infrastructure sizing is obtained by solving
\begin{equation} \label{eq:argmin-lcc}
\theta^{\star}
=
\arg\min_{\theta \in \Theta_{\text{feas}}}
LCC(\theta).
\end{equation}

Importantly, reinforcement learning is not used to directly optimize $\theta$. Instead, RL serves as a mechanism for estimating the operational cost component of the objective under optimal control. This decoupling ensures that the sizing optimization remains transparent, interpretable, and aligned with standard life-cycle cost analysis practices in energy systems engineering.

\section{Experimental Results}\label{sec:simulations}


This section evaluates the proposed reinforcement learning framework as an operational cost estimator for the HVAC co-design problem. The evaluation focuses on two questions:
\begin{enumerate}
    \item Operational performance: For a fixed chiller–TES sizing configuration, how does the learned DQN policy compare against baseline operational policies in terms of electricity cost and loss-of-cooling-load?
    \item Design implications: How do these operational outcomes translate into feasible sizing configurations and life-cycle cost trade-offs?
\end{enumerate}
All experiments are conducted using historical hourly cooling demand and electricity price traces spanning one full year ($T=8760$ hours) to evaluate the performance of various agents. These data traces were obtained from a high rise residential building in Hyderabad, India.

\subsection{Experimental Setup}

We consider a discrete set of infrastructure configurations $\theta
=\bigl(
C_{\mathrm{ch}},
E_{\max}
\bigr),$
spanning undersized to moderately oversized chiller capacities paired with varying TES capacities (Table~\ref{tab:performance-best}). For each configuration, we evaluate four operational policies:
\begin{itemize}
\item Greedy (Myopic) Policy,
\item TES-First Feasibility Policy (TFP),
\item DQN-based RL Policy,
\item Storage-Dominant Pessimistic Policy (SDPP).
\end{itemize}
For each $\theta$, a DQN RL policy is trained over 50 episodes, each corresponding to a full year of operation. The historical demand and price traces are identical across episodes, while the initial TES state of charge is randomized to improve coverage of storage states. After training, the learned policy is evaluated once over the full year and compared against baselines. 

\paragraph{Training with Limited Historical Data}

Although only a single year of demand and price data is available, randomizing the initial TES state across episodes induces diverse state trajectories during training. Since TES state directly couples decisions across time, this strategy allows the DQN to learn cost-to-go values across a wide range of operational contexts. Randomizing the initial TES state induces coverage over
the feasible storage state space and allows the Q-function to learn
cost-to-go values across a wide range of operating conditions, helping the model generalize better and prevent overfitting and reamained aligned with the outer-loop sizing objective.

\paragraph{Performance Metrics}

For each policy and configuration, we record: Annual electricity cost
$f_{\mathrm{elec}}(\theta),$
Cumulative loss-of-cooling-load and number of loss-of-load events, TES energy throughput. Loss-of-load is treated as a hard feasibility violation. Configurations with nonzero loss-of-load are excluded from life-cycle cost analysis.

\begin{table*}
    \centering
    \caption{Comparison of greedy, RL, TFP, and SDPP policies over a year-long trace for varying chiller and TES configurations. $\pi_{\text{SDPP}}$ being a theoretical extreme which conserves stored energy at all times yielded 0 loss-of-load for valid sizing configurations}
    \begin{tabular}{|c|c|c|c|c|c|c|c|c|c|c|c|c|c|}
        \hline
        Capacity & \multicolumn{4}{c|}{Yearly cost $f_{elec}(\cdot)$ in rupees} & \multicolumn{3}{c|}{Total loss of LoL} & \multicolumn{3}{c|}{\# LOL incidents} & \multicolumn{3}{c|}{TES throughput}\\
        \cline{2-14}
        ($C_\text{ch}$, $E_{max}$) & $\pi_{\text{grd}}$ & $\pi_{\text{RL}}$ & $\pi_{\text{TFP}}$ & $\pi_{\text{SDPP}}$* & $\pi_{\text{grd}}$ & $\pi_{\text{RL}}$ & $\pi_{\text{TFP}}$ & $\pi_{\text{grd}}$ & $\pi_{\text{RL}}$ & $\pi_{\text{TFP}}$ & $\pi_{\text{grd}}$ & $\pi_{\text{RL}}$ & $\pi_{\text{TFP}}$\\
        \hline
        (400, 4000) & 1,742,868 & 1,651,089 & 1,524,884 & 6,242,156 & 118,072 & 0 & 111,227 & 643 & 0 & 709 & 394,673 & 548,369 & 475,029 \\
        (400, 5000) & 1,742,337 & 1,652,475 & 1,524,356 & 6,222,899 & 117,969 & 0 & 111,108 & 640 & 0 & 711 & 394,696 & 550,217 & 474,902 \\
        (500, 2500) & 1,861,682 & 1,681,436 & 1,608,746 & 8,413,517 & 59,486 & 0 & 55,859 & 409 & 0 & 449 & 476,702 & 605,969 & 569,664 \\
        (500, 3000) & 1,860,677 & 1,679,208 & 1,607,591 & 8,401,260 & 59,508 & 0 & 55,815 & 409 & 0 & 447 & 476,469 & 607,176 & 569,744 \\
        (600, 2000) & 1,947,432 & 1,691,782 & 1,666,238 & 10,486,170 & 23,800 & 32 & 18,390 & 225 & 1 & 252 & 555,517 & 660,683 & 649,153 \\
        (600, 2500) & 1,945,210 & 1,691,881 & 1,665,545 & 10,482,488 & 23,764 & 0 & 18,371 & 223 & 0 & 245 & 555,663 & 661,560 & 649,464 \\
        (600, 3000) & 1,944,055 & 1,693,245 & 1,665,545 & 10,482,582 & 23,763 & 0 & 18,362 & 223 & 0 & 253 & 556,145 & 659,776 & 649,874 \\
        (700, 1500) & 1,997,586 & 1,694,901 & 1,689,472 & 12,500,442 & 6,240 & 0 & 3,413 & 104 & 0 & 136 & 620,722 & 718,105 & 717,520 \\
        (700, 1800) & 1,998,451 & 1,695,479 & 1,689,587 & 12,503,031 & 6,218 & 0 & 3,411 & 101 & 0 & 144 & 620,813 & 717,256 & 717,797 \\
        (800, 1500) & 2,038,192 & 1,706,271 & 1,699,538 & 14,303,084 & 263 & 0 & 829 & 21 & 0 & 91 & 676,109 & 780,966 & 780,479 \\
        (800, 1800) & 2,037,166 & 1,702,970 & 1,698,350 & 14,313,001 & 261 & 0 & 837 & 20 & 0 & 93 & 676,384 & 783,455 & 780,842 \\
        \hline
    \end{tabular}
    \label{tab:performance-best}
\end{table*}

\begin{table*}
    \centering
    \caption{Life cycle cost analysis with RL operational policy across representative chiller–TES sizing. The CAPEX costs are Rs.6410 (chiller) and Rs.1500 (TES) per $kWh_{th}$ storage capacity \cite{LUERSSEN2019640}.}
    \begin{tabular}{|c|c|c|c|c|}
        \hline
        Capacity $(kWh_{th})$ & 
        CAPEX costs & 
        Yearly electricity & 
        OPEX costs & 
        Life Cycle Cost \\
        $\theta=$ ($C_\text{ch}, \,E_{max}$) & (chiller + TES) & cost $f_{elec}(\theta)$ &  &  \\
        \hline
        (400, 4000) & 8,564,000 & 1,651,089 & 26,966,246 & 35,530,246 \\
        (400, 5000) & 10,064,000 & 1,652,475 & 27,727,844 & 37,791,844 \\
        (500, 2500) & 6,955,000 & 1,681,436 & 26,587,487 & 33,542,487 \\
        (500, 3000) & 7,705,000 & 1,679,208 & 26,928,082 & 34,633,082 \\
        (600, 2500) & 7,596,000 & 1,691,881 & 27,048,571 & 34,644,571 \\
        (600, 3000) & 8,346,000 & 1,693,245 & 27,438,606 & 35,784,606 \\
        \textbf{(700, 1500)} & 6,737,000 & 1,694,901 & 26,664,922 & \textbf{33,401,922} \\
        (700, 1800) & 7,187,000 & 1,695,479 & 26,895,628 & 34,082,628 \\
        (800, 1500) & 7,378,000 & 1,706,271 & 27,138,726 & 34,516,726 \\
        (800, 1800) & 7,828,000 & 1,702,970 & 27,316,053 & 35,144,053 \\
        \hline
    \end{tabular}
    \label{tab:life-cycle-cost}
\end{table*}

\subsection{Results}
\paragraph{Loss-of-Cooling-Load and Feasibility}

Table~\ref{tab:performance-best} reports the performance of the greedy, RL, TES-First Feasibility Policy (TFP), and SDPP over a year-long trace for a range of chiller--TES sizing configurations. A clear feasibility boundary emerges with respect to chiller capacity when loss of cooling load under the operating policy is considered. While SDPP, being a conservative policy that always operate chiller at max part-load-ratio, results is zero loss of cooling load across any sizing configuration\footnote{SDPP incur zero loss of load in all sizing confgurations and hence not reported in the Table~\ref{tab:performance-best}}, substantial loss-of-cooling-load is observed under both the greedy and TFP policies, even when paired with large TES capacities. This indicates that, in these configurations, peak cooling demand exceeds the combined instantaneous supply capability of the chiller and admissible TES discharge.

In contrast, the RL policy was almost always able to achieve zero loss-of-cooling-load clearly indicating its superiority to operate in meeting cooling load reliably and cost effectively. This demonstrates that, with a minimum level of chiller capacity, anticipatory operation enabled by RL can fully satisfy cooling demand, whereas myopic or rule-based policies may still fail.

These results confirm that feasibility is jointly determined by infrastructure sizing and operational intelligence: while infeasibility at very small chiller capacities is unavoidable for any policy, the RL policy strictly enlarges the set of sizing configurations that achieve zero loss-of-cooling-load.




\paragraph{Operational Cost Comparison}

In addition to achieving zero loss-of-cooling-load, Table~\ref{tab:performance-best} shows that RL consistently yields lower annual electricity cost than the other baseline policies (barring few exceptions). The cost reduction is most pronounced for moderate chiller capacities paired with nontrivial TES capacities (e.g., $(500,3000)$, $(600,2500)$, $(700,1800)$), where the RL policy reduces annual electricity cost by approximately 8--12\% relative to greedy operation. 

The greedy policy, which always selects the minimum feasible part-load ratio, systematically underutilizes TES and fails to anticipate future price spikes, resulting in higher electricity costs even when loss-of-load is avoided. In contrast, the RL policy actively exploits TES, as evidenced by substantially higher TES throughput across all feasible configurations. This increased throughput reflects strategic charging during low-price periods and discharge during high-price periods, consistent with the implicit marginal valuation of storage described in Section~\ref{sec:all_policies}. Note that SDPP, being a theoretical extreme that always tries to keep storage full at all possible times, naturally yields highest electricity costs as shown in Table~\ref{tab:performance-best}.

\paragraph{Implications for Life-Cycle Cost Minimization}

The operational results directly help with the sizing optimization in Section~\ref{sec:resource_sizing}.
Since the RL policy dominates the baseline policies in terms of feasibility (achieving zero loss of cooling load across a broader set of sizing configurations) the operational cost attained by the RL agent for each feasible configuration provides a meaningful basis for sizing optimization. Among all configurations that meet the cooling demand reliably over the one-year horizon, the configuration with the lowest RL operational cost naturally emerges as the optimal cooling resource sizing.

Building on this observation, Table~\ref{tab:life-cycle-cost} reports the life-cycle cost analysis for representative sizing configurations operated exclusively under the RL policy. By restricting attention to RL-feasible designs, the table isolates the trade-off between capital expenditure and RL-enabled operational cost. 
Among the RL-feasible designs considered in Table~\ref{tab:life-cycle-cost}, the configuration $(C_{\mathrm{ch}}, E_{\max}) = (700~\mathrm{kWh}_{\mathrm{th}}, 1500~\mathrm{kWh}_{\mathrm{th}})$ achieves the minimum life-cycle cost, identifying it as the cost-optimal cooling infrastructure sizing under RL-based operation.
 The results show that the cost-optimal design does not correspond to the largest chiller, but instead to a balanced configuration in which operational intelligence substitutes for capital-intensive overprovisioning.



\section{Conclusion}
Through this study, we successfully co-design the required infrastructure size for the HVAC system. We demonstrate the clear economic benefit of using the thermal energy storage, both in reducing the initial capital investment, as well as reduced operational cost by intelligently scheduling energy through the TES. The optimal operation schedule problem is posed as a Markov Decision Process, solved through Reinforcement Learning. We demonstrate the non-triviality of the optimal policy design through the marginal value of storage. Through a Deep Q Network, trained over a reward function that incentivizes low electricity cost and heavily penalizes loss-of-load, we are able to learn the optimal operation policy that minimizes the operational electricity cost while avoiding instances of loss of cooling load. We evaluate this policy through comparisons with other baseline policies that vary in their behaviour with respect to the marginal value of energy storage. Our results clearly show the superior performance of the RL agent, both in terms of yearly operational cost, and loss of load with the baseline policies. Overall, the experimental findings validate the use of reinforcement learning as a computational tool for estimating optimal operational cost of meeting cooling load and its use for planning optimal cooling infrastructure.  

\bibliographystyle{ACM-Reference-Format}
\bibliography{acmart}










\end{document}